\documentclass[letter,12pt]{article}
\usepackage{amsmath, amssymb}
\usepackage{graphicx}

\title{A new mean field approach to finite spin systems}

\author{Gabriel Gil and Augusto Gonzalez\\
Institute of Cybernetics, Mathematics and Physics, Havana, Cuba}

\begin{document}

\maketitle

\begin{abstract}
It is shown that a spin system is equivalent to a set of constrained harmonic oscillators.
For finite, but large, systems, a continuous approximation to the density of states can
be used, and the oscillator frequencies can be exactly computed. In the phase transition, 
the effective frequency of the lowest mode passes through zero, that is, it becomes an inverted 
oscillator. In the small oscillations regime, the oscillators can be treated as independent
and the thermodynamic magnitudes can be computed. We show explicit calculations in a disordered,
frustrated, high coordination number Blume Capel model with $6\times 10^4$ spins.
\end{abstract}

\section{Introduction}
Spin systems have attracted a lot of attention recently. Experimental studies on glassy 
systems \cite{SG.Exp,SG.Exp2}, and new theoretical ideas \cite{Parisi.Book} have moved ahead the
theoretical understanding in this research area. Although the goal is the thermodynamic limit,
concepts are usually tested against numerical Monte-Carlo calculations on finite sytems
\cite{MC,MC2,MC3}, which are afterward extrapolated to the infinite limit.

The finite system itself is the hardest from the theoretical point of view. Phases, transitions 
and other thermodynamic concepts can not be used. On the other hand, exact calculations can be 
performed only on very small systems. The number of configurations in a Ising lattice with
100 spins, for example, is around $10^{30}$, a number beyond reach. Monte-Carlo calculations,
on the other hand, suffer from metastability effects in glassy systems due to the existence
of many local quasi-degenerated pure states \cite{Parisi.Book,MC,MC2,MC3}.

The reason mentioned above explains why to get a quasi-exact approximation for large finite
spin systems is a highly desirable goal. In the present paper, we provide a kind of mean-field
approximation \cite{MF,MF2,MF3}. The spin system is shown to be equivalent to a set of constrained 
oscillators. When the number of spins is high enough, we may use a continuous approximation for
the density of states, and the effective frequencies of these oscillators can be exactly computed.
The phase transition in the spin system is apparent as a change of sign in the frequency of the 
lowest mode. That is, it becomes an inverted oscillator. In the regime of small oscillations, the 
oscillators are decoupled and the thermodynamic magnitudes can be computed. 

For illustrative purposes, in the paper we perform calculations on a $S=1$ Blume-Capel model (BCM)
\cite{BCM,BCM2,BCM3} with Hamiltonian:

\begin{equation}
H=-\frac{1}{2\sqrt{N}}\sum_{i,j}K_{ij}S_iS_j+D\sum_iS_i^2.
\label{eq1}
\end{equation}

\noindent
The BCM is usually employed to describe the magnetic properties of systems with randomly distributed 
impurities, but for us it has a biological motivation. The spin variables, $S_i$, take values -1, 0 and
1. The parameter $D$ is a kind of chemical potential, allowing to control the number of ``excited''
spins, i.e. spins with $|S_i|=1$.

We shall study a system with $N=6\times 10^4$ spins, a number which is a challenge for current
Monte-Carlo calculations. The spin-spin couplings, $K_{ij}$ is described by a symmetric matrix that
take random integer values in $\{-1,0,1\}$. Of course, $K_{ii}=0$. The number of ferromagnetic couplings
(+1) is roughly equal to the number of anti-ferromagnetic ones (-1). And, overall, the number of non-zero
$K_{ij}$ is roughly $0.5\%$ of $N^2$. It means that, in average, each spin interacts with $300$
neighbors, and mean field approximations should work.

Thus, our spin model exhibits disorder, frustration and high coordination numbers. In addition, we
will not average over disorder, that is, shall perform calculations for a given realization of the 
$K_{ij}$.

\section{The spin system as a set of constrained oscillators}
The method to be applied is based on the idea that the matrix $K_{ij}$ can be decomposed in terms
of its eigensystem:

\begin{equation}
-\frac{1}{2\sqrt{N}}K_{ij}=\sum_{n}\lambda_n u_{ni}u_{nj}.
\label{eq2}
\end{equation}

\noindent
The $\lambda_n$ are the (ordered) eigenvalues, and $u_n$ the corresponding eigenvectors. The 
eigenvalues are real and satisfy the equation $0={\rm Tr}~K=\sum_n \lambda_n$.

\begin{figure}
\includegraphics[width=0.8 \linewidth]{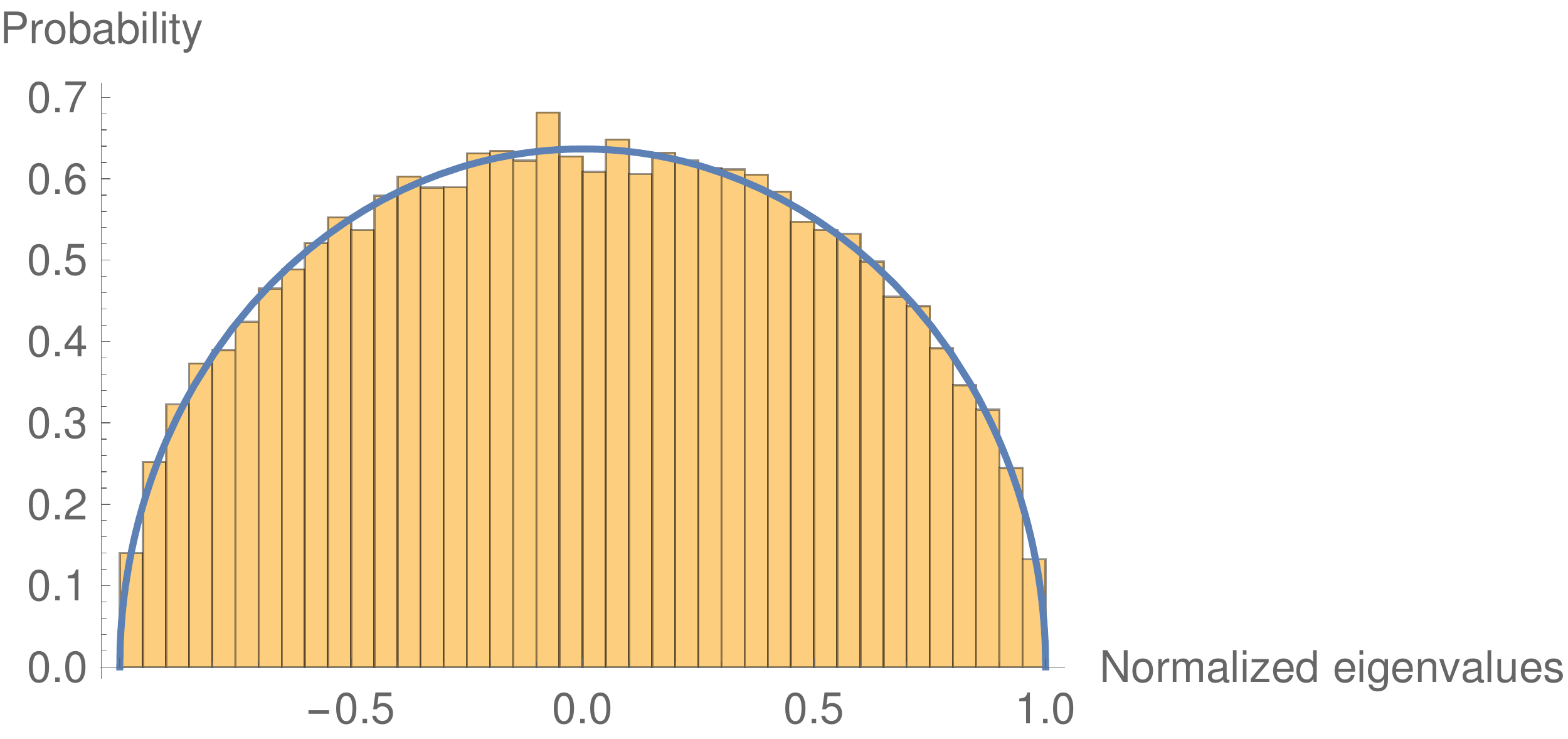}
\caption{The distribution of normalized eigenvalues for the matrix $-K_{ij}/(2\sqrt{N})$.}
\label{fig1} 
\end{figure}

For the system under study, the normalized eigenvalues, $y_n=\lambda_n/|\lambda_1|$, are
distributed according to Wigner semi-circle law \cite{WSL,WSL2}, 
$2 \sqrt{1-y^2}/\pi$, as it is apparent in Fig. \ref{fig1}.

The lowest eigenvalue, $\lambda_1=-0.0707$ is roughly minus the square root of the sparseness index. 
By using the decomposition (\ref{eq2}), the Hamiltonian is written as a sum of oscillators:

\begin{equation}
H = \sum_n (\lambda_n+D)x_n^2,
\label{eq3}
\end{equation}

\noindent
where 

\begin{equation}
x_n = \sum_i S_i u_{ni}, 
\label{eq4}
\end{equation}

\noindent
is the projection of the lattice spin configuration over the eigenvector $u_n$. In our
model, with equilibrated ferro- and anti-ferro couplings, the $u_n$ vectors corresponding
to the lowest $\lambda_n$ have large ${\rm Tr}|u_n|$, but relatively small ${\rm Tr} (u_n)$.
That is, the typical configurations have large $\langle S^2\rangle$, but small magnetizations 
$\langle S\rangle$. This may change if the proportion of ferro- to anti-ferro bonds changes.

The variables $x_n$ are constrained to take discrete values in the interval:

\begin{equation}
 -{\rm Tr} |u_n| \le x_n \le {\rm Tr} |u_n|.
\label{eq5}
\end{equation} 

\noindent
In the model, the average value of ${\rm Tr} |u_n|$ is $0.8 \sqrt{N}$, that is around 196.
In addition, the $x_n$ variables are not independent. Indeed, for a given spin configuration, 
we may write:

\begin{equation}
 \vec S = \sum_n (\vec S \cdot u_n) u_n =\sum_n x_n u_n,
\label{eq6}
\end{equation}

\noindent
and then, using orthogonality of the $u_n$:

\begin{equation}
 \label{eq7}
 \vec S\cdot\vec S=\sum_n x_n^2.
\end{equation}

\noindent
The l.h.s. of Eq. (\ref{eq7}) takes values between 0 and $N$.

The partition function in our model can thus be written:

\begin{equation}
 \label{eq8}
 Z= \left(\prod_n Z_n\right) \Theta\left(N-\sum_n x_n^2\right),
\end{equation}

\noindent 
where

\begin{equation}
 \label{eq9}
 Z_n= \sum_{x_n}e^{-(\lambda_n+D)x_n^2/T},
\end{equation}

\noindent 
and $\Theta$ is the step (Heavyside) function: $\Theta(y)=1$ when $y>0$,
and $\Theta(y)=0$ for $y<0$. 

\section{The continuous approximation to $Z_n$}
As mentioned, the $x_n$ take discrete values in the interval given by Eq. (\ref{eq5}). 
There is a single configuration for which $x_n={\rm Tr}|u_n|$. In that case 
$S_i={\rm Sign}(u_{ni})$. But there are plenty of configurations with $x_n\approx 0$.
In the large $N$ limit, the projections of spin configurations are quasi-continuous
and distributed according to the density:

\begin{equation}
 \label{eq10}
 \rho(x_n)= \sqrt{\frac{3}{4\pi}}e^{-3 x_n^2/4}.
\end{equation}

\noindent
In other words, the projection of $\vec S$ over $u_n$ behaves as a single $S=1$ spin.

The mode partition function $Z_n$ can be written:

\begin{equation}
 \label{eq11}
 Z_n\approx \int_{-{\rm Tr}|u_n|}^{{\rm Tr}|u_n|} e^{-\omega_n x_n^2}~{\rm d}x_n,
\end{equation}

\noindent 
where the frequencies $\omega_n$ are given by:

\begin{equation}
 \label{eq12}
 \omega_n= \frac{\lambda_n+D}{T}+\frac{3}{4}.
\end{equation}

This quasi-continuous approximation is essential for the results of the paper. It can be seen to be equivalent to
a mean field approximation.
 
\section{The phase transition}
The effective frequencies given in Eq. (\ref{eq12}) have two contributions. The first coming from the 
system parameters, and a second ``entropic'' contribution coming from the number of available states.
Even if $(\lambda_n+D)<0$ it should overcome the second contribution in order to $\omega_n$ to become negative.

\begin{figure}
\includegraphics[width=0.8 \linewidth]{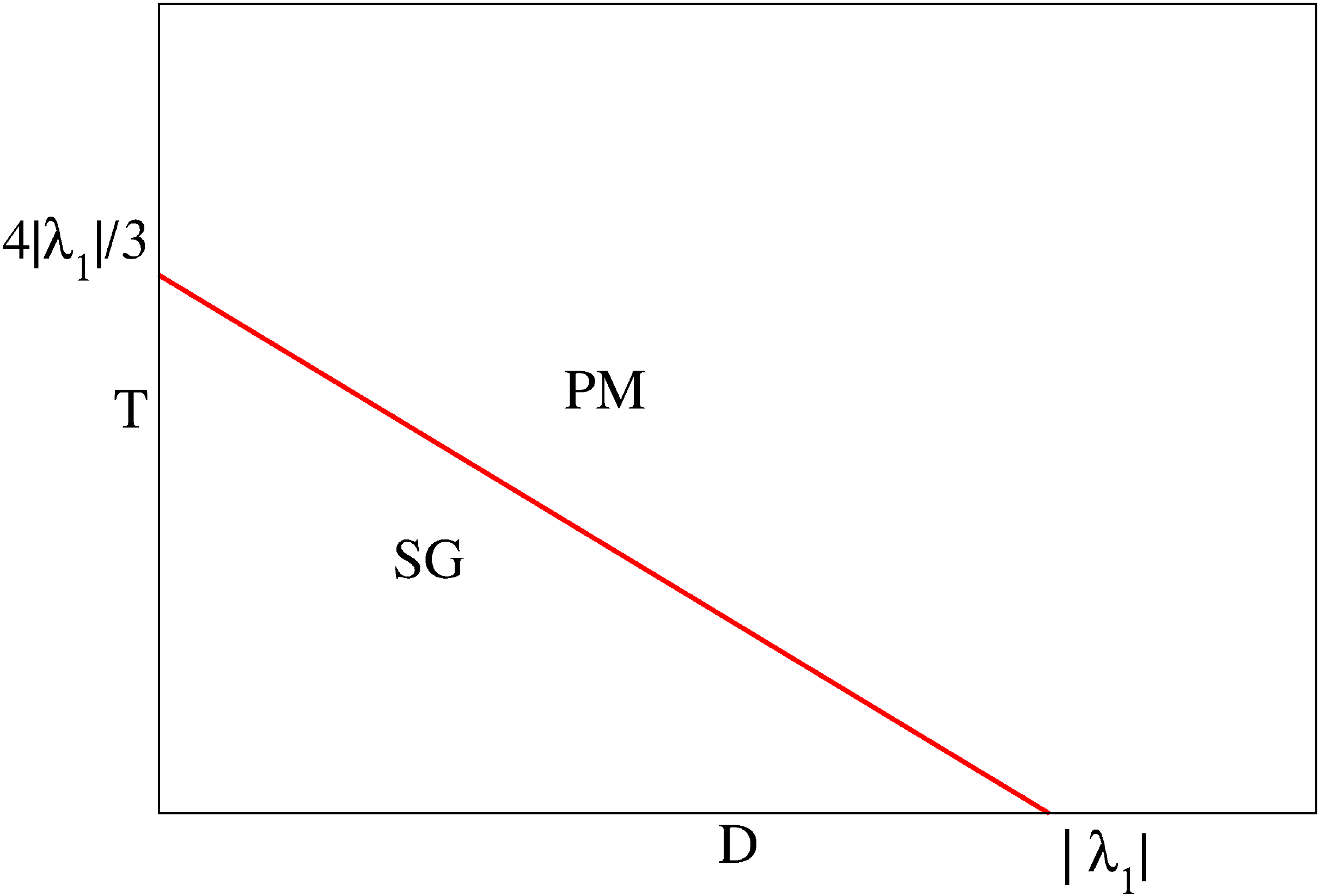}
\caption{Phase diagram of the studied BCM.}
\label{fig2}
\end{figure}

And, indeed, there is a change of behavior when $\omega_n$ changes sign. In particular $\omega_1$, 
corresponding to the lowest eigenvalue $\lambda_1$. For $D>|\lambda_1|$, $\omega_1$ is always greater
than zero. When $0\le D<|\lambda_1|$, one can define a critical temperature $T_c=4(|\lambda_1|-D)/3$
signaling the transition from the paramagnetic state (PM) to the spin glass (SG). The phase diagram
of the studied BCM is shown in Fig. \ref{fig2}. 

It may be verified, in Ising or other systems in low dimensions, that the critical temperature deduced from  
the change of sign in $\omega_1$ is related to a mean field approximation. In our model, with relatively
high coordination numbers, this approximation should be nearly exact. 

\section{Small oscillations}
The regime of small oscillations is characterized by $\sum_n x_n^2<<N$. In the model, this regime can 
be reached with relatively high values of $D$ or $T$. The $\Theta$ function in this case evaluates to 1, 
and the mode partition functions, $Z_n$ can be independently computed.

\subsection{The mean value $\langle x_n^2\rangle$}
In the calculation of $\langle x_n^2\rangle$ we shall distinguish the cases $\omega_n>0$ and $\omega_n<0$.

\begin{equation}
 \label{eq13}
 \omega_n>0:~~~\langle x_n^2\rangle= \frac{1}{2\omega_n}
  \left(
  1-\frac{ \sqrt{\omega_n}~{\rm Tr} |u_n|~e^{-\omega_n({\rm Tr}|u_n|)^2} }
  { \sqrt{\pi}~{\rm Erf}(\sqrt{\omega_n}~{\rm Tr}|u_n|)/2 }
  \right),
\end{equation}

\noindent where ${\rm Erf}(y)=2\int_0^y \exp(-t^2){\rm d}t/\sqrt{\pi}$ is the error function.
The limiting values are:

\begin{equation}
 \label{eq14}
 \langle x_n^2\rangle=0,~~T\to 0~~{\rm and}~~D>|\lambda_n|,
\end{equation}

\begin{equation}
 \label{eq15}
 \langle x_n^2\rangle=2/3,~~T\to \infty,
\end{equation}

\begin{equation}
 \label{eq16}
 \langle x_n^2\rangle=({\rm Tr}|u_n|)^2/3,~~\omega_n\to 0.
\end{equation}

On the other hand,

\begin{equation}
 \label{eq17}
 \omega_n < 0:~~~ \langle x_n^2\rangle = \frac{1}{2|\omega_n|}
  \left(
  -1+\frac{ \sqrt{|\omega_n|}~{\rm Tr} |u_n| }
  { {\rm DawsonF}(\sqrt{|\omega_n|}~{\rm Tr}|u_n|) }
  \right),
\end{equation}

\noindent where ${\rm DawsonF}(y)=e^{-y^2} \int_0^y e^{t^2}{\rm d}t$ is the Dawson function.
The limiting values are:

\begin{equation}
 \label{eq18}
 \langle x_n^2\rangle=({\rm Tr}|u_n|)^2,~~T\to 0,
\end{equation}

\begin{equation}
 \label{eq19}
 \langle x_n^2\rangle=({\rm Tr}|u_n|)^2/3,~~\omega_n\to 0.
\end{equation}

\begin{figure}
\includegraphics[width=0.8 \linewidth]{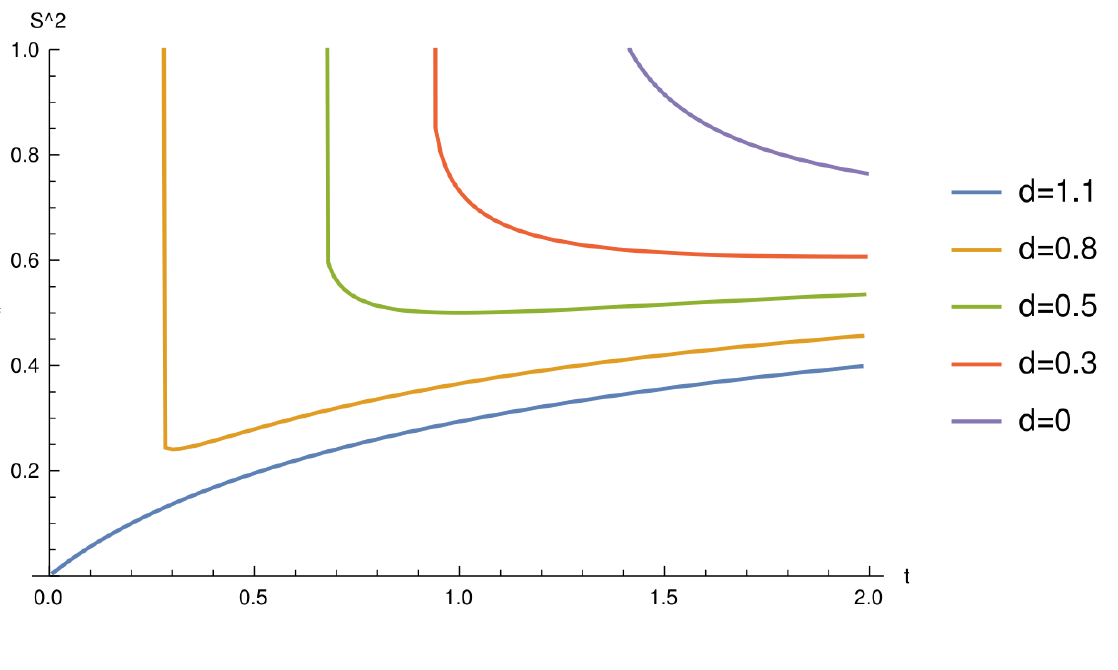}
\caption{$\langle S^2\rangle$ as a function of $t=T/|\lambda_1|$ and $d=D/|\lambda_1|$ in the BCM.}
\label{fig3}
\end{figure}

\subsection{The mean value of $S^2$}
The mean value of $S^2$ is defined as:

\begin{equation}
 \label{eq20}
 \langle S^2\rangle=\frac{1}{N} \left\langle \sum_n x_n^2\right\rangle.
\end{equation}

In the small oscillations regime, the mode averages can be independently computed. 
The results are shown in Fig. \ref{fig3}.

All of the curves tend to 2/3 in the high-$T$ limit and exhibit an abrupt rise as
$t$ approaches $t_c$. The transition for $d\lesssim 0.7$ is beyond the plot range 
and beyond the validity range of the small oscillations approximation because, according 
to Eq. (\ref{eq16}), near the transition point the lowest modes make contributions 
$({\rm Tr}|u_n|)^2/3\approx 0.2~N$ to $\langle S^2\rangle$.

Note that, for high enough values of $d$, there are points inside the PM phase at which 
${\rm d}\langle S^2\rangle/{\rm d}T=0$. They are related to the mentioned rise of $\langle S^2\rangle$ 
as $T$ approaches $T_c$.

\subsection{The mean energy}
The mean energy per spin is defined as:

\begin{equation}
 \label{eq21}
 \langle E\rangle=\frac{1}{N} \left\langle \sum_n \lambda_n x_n^2\right\rangle.
\end{equation}

In the small oscillations regime it can be straightforwardly computed from the 
previous results. For temperatures greater than $T_c$, all the $\omega_n$
are greater than zero and:

\begin{equation}
 \label{eq22}
 \langle E\rangle\approx\frac{1}{N} \sum_n \frac{2}{3}\lambda_n\to 0,~~T\to\infty.
\end{equation}

\begin{figure}
\includegraphics[width=0.8 \linewidth]{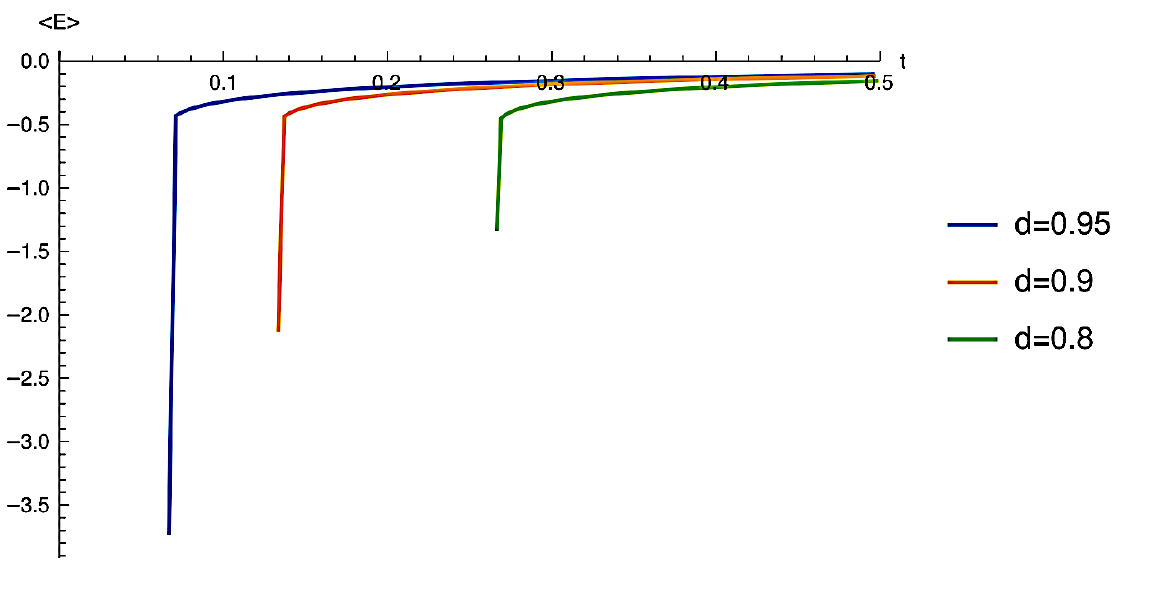}
\caption{$\langle E\rangle$ as a function of the reduced $t$ and $d$ in the BCM.
Each curve starts at the corresponding $t_c$ and ends at $t=0.5$.}
\label{fig4}
\end{figure}

The results are shown in Fig. \ref{fig4} for $d>0.8$ and $t\ge t_c$, the range in which the small oscillations
approximation holds. Notice the very high slope at $t_c$.

\section{Discussion}
In the paper, the interaction matrix of a spin system is diagonalized in order to represent the
system as a set of constrained oscillators. For large enough lattices, a quasi-continuous
approximation to the projections of the spin configurations over the eigenvectors allows the
exact computation of oscillator frequencies, leading to a new mean field theory for the 
spin system. In quality of illustration, the method is applied to obtain observables in a
disordered, frustrated BCM with 60,000 spins.

In the small oscillations regime, the oscillators are decoupled. This approximation holds
at large enough temperatures or high values of the $D$ parameter. Below the transition 
temperature, the constrained sum, Eq. (\ref{eq8}), shall be used. We stress that this 
expression is very well suited for Monte Carlo evaluations. In this way, a mean field
Monte Carlo scheme may emerge. Work along this direction is in progress. 

\vspace{.5cm}
{\bf Acknowledgements}
Authors acknowledge the Cuban Agency for Nuclear Energy and Advanced Technologies (AENTA) 
and the the Office of External Activities of the Abdus Salam Centre for Theoretical Physics (ICTP) for support. 
The authors are grateful to C. Chatelain for suggestions and criticism.


\begin{thebibliography}{50}
\bibitem{SG.Exp} K. Binder, A.P. Young. Spin glasses: Experimental facts, theoretical concepts,
and open questions. Reviews of Modern Physics, Vol. 58, (1986) 801.
\bibitem{SG.Exp2} J.A. Mydosh. Spin glasses: redux: an updated experimental/materials survey.
Rep. Prog. Phys. 78 (2015) 052501.
\bibitem{Parisi.Book} M. Mezard, G. Parisi, M.A. Virasoro. Spin glass theory and beyond. Lecture Notes 
in Physics Vol. 9. World Scientific. Singapore 1987.
\bibitem{MC} Robert H. Swendsen and Jian-Sheng Wang. Replica Monte Carlo Simulation of Spin-Glasses.
Phys. Rev. Lett. 57 (1986) 2607.
\bibitem{MC2} Koji Hukushima and Koji Nemoto. Exchange Monte Carlo method and application to spin
glass simulation. Journal of the Phys. Soc. of Japan 65 (1996) 1604-1608.
\bibitem{MC3} Helmut G. Katzgraber, Matteo Palassini and A. P. Young (2001). Monte Carlo Simulations of 
Spin Glasses at Low Temperatures. Phys. Rev. B 63, 184422.
\bibitem{MF} Tommaso Castellani and Andrea Cavagna.Spin-Glass Theory for Pedestrians.
J. Stat. Mech. (2005) P05012. DOI 10.1088/1742-5468/2005/05/P05012.
\bibitem{MF2} Francesco Zamponi (2014). Mean field theory of spin glasses. arXiv:1008.4844.
\bibitem{MF3} Leticia F. Cugliandolo. Course 7: Dynamics of Glassy Systems. 
In: Barrat, JL., Feigelman, M., Kurchan, J., Dalibard, J. (eds) 
Slow Relaxations and nonequilibrium dynamics in condensed matter. 
Les Houches-École d’Été de Physique Theorique, vol 77. Springer, Berlin, 
Heidelberg. https://doi.org/10.1007/978-3-540-44835-8\_7.
\bibitem{BCM} J.A. Plascak, J.G. Moreira and F.C. Barreto. Mean field solution of the general 
spin Blume-Capel model. Physics Letters A 173 (1993) 360-364.
\bibitem{BCM2} D. Peña Lara, J. E. Diosa, and C. A. Lozano. Blume-Capel spin-glass model 
for Fe-Mn-Al alloys. Phys. Rev. E 87, 032108 (2013).
\bibitem{BCM3} L. Leuzzi, M. Paoluzzi, and A. Crisanti. Random Blume-Capel model on a 
cubic lattice: First-order inverse freezing in a three-dimensional spin-glass system.
Phys. Rev. B 83, 014107 (2011).
\bibitem{WSL} Florent Benaych-Georges and Antti Knowles. Lectures on the local semicircle 
law for Wigner matrices. In Advanced topics in random matrices, 1--90, Panor. Synthèses, 53, 
Soc. Math. France, Paris, 2017.
\bibitem{WSL2} Laszlo Erdos, Benjamin Schlein and Horng-Tzer Yau. Local semicircle law 
and complete delocalization for Wigner random matrices. Commun. Math. Phys. 287, 
pages 641–655 (2009).
\end{thebibliography}
\end{document}